\documentstyle[aps,prl,twocolumn,psfig,draft,floats]{revtex}

\newcommand{\fig}[1]{Fig.~\ref{fig:#1}}

\newcommand{\exf}[2]{\mbox{$#1\!\times\! 10^{#2}$}}

\begin{document}

\wideabs{
\title{Viscous stabilization of 2D drainage displacements with trapping}
\author{Eyvind Aker,$^{1,2}$ Knut J\o rgen  M\aa l\o y,$^1$ and Alex Hansen$^2$}
\address{$^1$Department of Physics, University of Oslo, N-0316 Oslo, Norway}
\address{$^2$Department of Physics, Norwegian University of Science and 
Technology, N-7491 Trondheim, Norway}
\date{\today}
\maketitle

\begin{abstract}
We investigate the stabilization mechanisms due to viscous forces in
the invasion front during drainage displacement in two-dimensional
porous media using a network simulator. We find that in horizontal
displacement the capillary pressure difference between two different
points along the front varies almost linearly as function of height
separation in the direction of the displacement. The numerical result
supports arguments taking into account the loopless displacement
pattern where nonwetting fluid flow in separate strands (paths). As a
consequence, we show that existing theories developed for viscous
stabilization, are not compatible with drainage when loopless strands
dominate the displacement process.
\end{abstract}

 } 

Immiscible displacement of one fluid by another fluid in porous media
generates front structures and patterns ranging from compact to
ramified and fractal~\cite{Maloy85_Chen-Wilk85,Len88,Cieplak88}.
When a nonwetting fluid displaces a wetting fluid (drainage) at low
injection rate, the nonwetting fluid generates a pattern of fractal dimension
equal to the cluster formed by invasion
percolation~\cite{Guyon78_Koplik82_Wilk83_Len85}.
The displacement is controlled solely by the capillary pressure,
that is the pressure difference between the two fluids across a pore
meniscus. At high injection rate and when the viscosity of the
nonwetting fluid is higher or equal to the viscosity of the wetting 
fluid, the width of the displacement front stabilizes and a more
compact pattern is generated~\cite{Len88,Frette97}

\begin{figure}
\begin{center}
\mbox{\psfig{figure=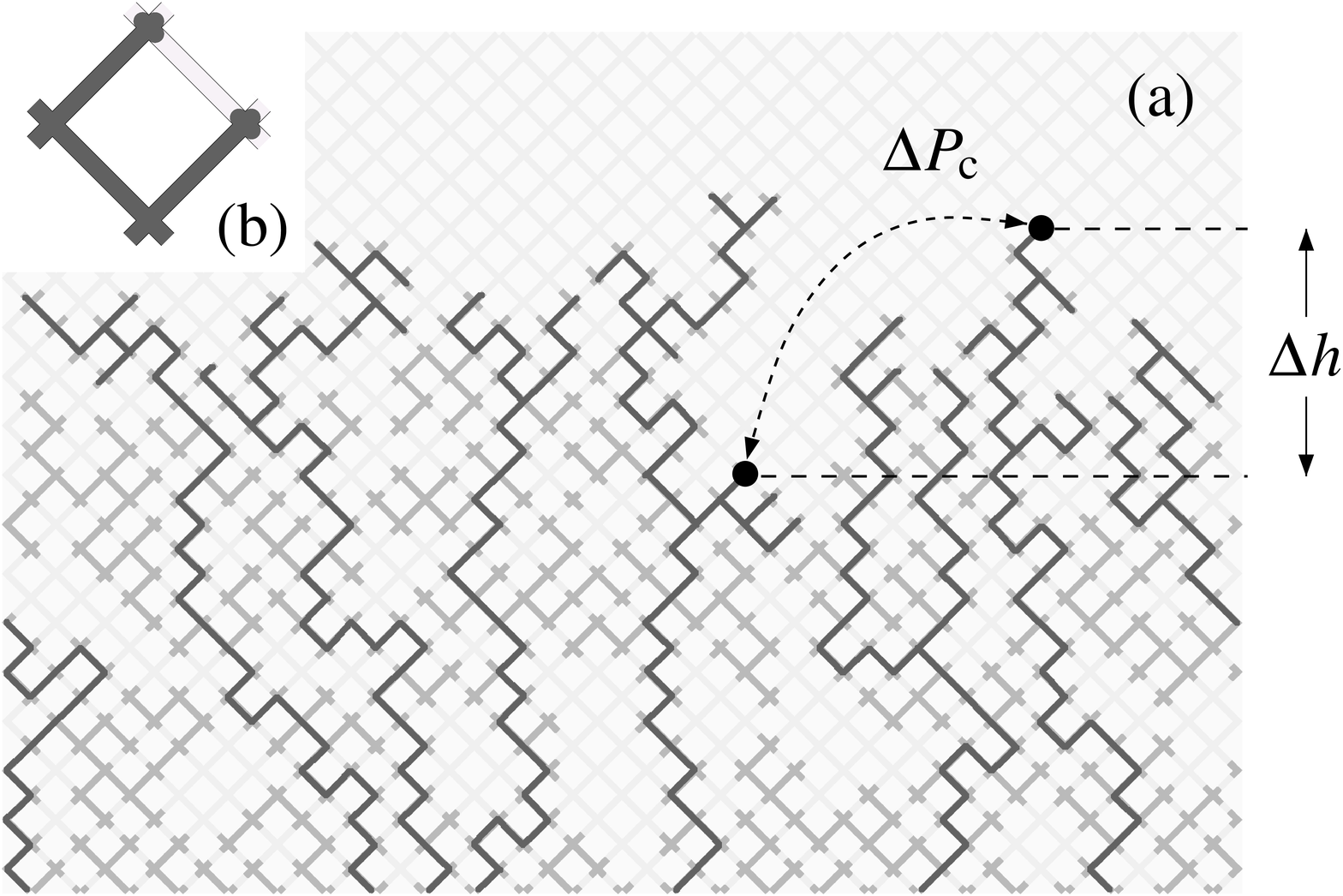,width=8cm}}
\caption{(a) The displacement pattern of a simulation on a lattice of
$25\times 35$ nodes. Nonwetting fluid (dark grey and black) is
injected from below and wetting fluid (light grey) flows out along the
top row.  In the figure, $\Delta P_c$ is the capillary pressure
difference between a meniscus at the bottom filled dot and a meniscus
at the topmost filled dot, separated a height $\Delta h$
in the direction of the displacement. The black tubes
indicate strands containing no loops, where nonwetting fluid flow. The
dark grey tubes connecting to the strands, are dead ends where
nonwetting fluid cannot flow because of trapped wetting fluid. (b) A
tube filled with wetting fluid and surrounded on both sides by
nonwetting fluid is trapped.}
\label{fig:model}
\end{center}
\end{figure}

\begin{figure}
\begin{center}
\mbox{\psfig{figure=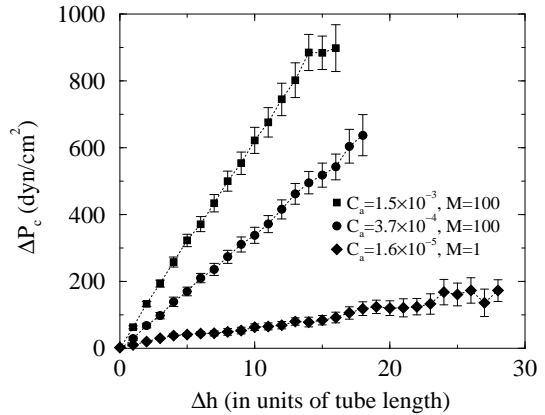,width=7cm}}
\caption{$\Delta P_c$ as function of $\Delta h$ for high and
intermediate $C_a$ with $M=100$ on a lattice of $25\times 35$ nodes,
and for low $C_a$ with $M=1$ on a lattice of $40\times 60$ nodes.}
\label{fig:pc-m100}
\end{center}
\end{figure}

\begin{figure}
\begin{center}
\mbox{\psfig{figure=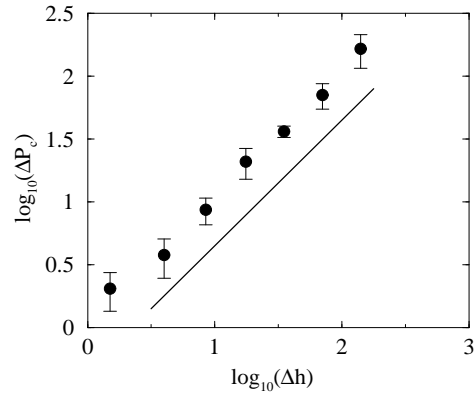,width=6.2cm}}
\caption{$\log_{10}(\Delta P_c)$ as function of $\log_{10}(\Delta h)$
for drainage simulations initiated on IP patterns at
$C_a=\exf{9.5}{-5}$ and $M=100$. The result is averaged over four
different runs and the error bars denote the standard error in the
mean. The slope of the solid line is 1.0.}
\label{fig:pcd-ip}
\end{center}
\end{figure}

The purpose of the present letter is to investigate the stabilization
mechanisms of the front due to viscous forces.To study the
stabilization mechanisms we consider two-dimensional (2D) horizontal
drainage at different injection rates. Since the displacement is
performed within the plane we neglect gravity. We present simulations
where we have calculated the capillary pressure difference $\Delta
P_c$ between two different pore menisci along the front separated a
height $\Delta h$ in the direction of the displacement
[\fig{model}(a)].  The simulations are based on a network model that
properly describes the dynamics of the fluid-fluid displacement as
well as the capillary and viscous pressure
buildup~\cite{Aker98-1,Aker98-2}. Simulations show that for a wide
range of injection rates and different fluid viscosities $\Delta P_c$
varies almost linearly with $\Delta h$ (Figs.~\ref{fig:pc-m100}
and~\ref{fig:pcd-ip}). Assuming a power law behavior $\Delta
P_c\propto \Delta h^\kappa$ we find $\kappa=1.0\pm 0.1$. This is a
surprising result because the viscous force field that stabilizes the
front, is non homogeneous due to trapping of wetting fluid behind the
front and to the fractal behavior of the front structure.

Based on the observation that the displacement structures are
characterized by loopless strands of nonwetting fluid
[\fig{model}(a)], we also present arguments being supported by our
numerical findings. We conjecture that the arguments might affect the
behavior of the front width $w_s$ as function of the capillary number
$C_a$. Here $C_a$ denotes the ratio between viscous and capillary
forces and in the following $C_a\equiv Q\mu_{nw}/\Sigma\gamma$, where
$Q$ is the injection rate, $\Sigma$ is the cross section of the inlet,
and $\mu_{nw}$ is the viscosity of the nonwetting phase.

In the literature~\cite{Wilk86,Len89,Blunt92,Xu98} there
has been suggested slightly different scaling behavior of $w_s$ as
function of $C_a$ and a general consensus has not yet been
reached. However, none of them consider the evidence observed here
that the displacement patterns are loopless and that nonwetting fluid
only flows in strands to displace wetting fluid.  As a consequence, we
show that earlier proposed theories~\cite{Wilk86,Len89,Blunt92,Xu98}
can not be used to describe drainage when loopless nonwetting strands
dominate the displacements.

Before we present the numerical results and the theoretical evidence,
we briefly introduce the network model. The model porous medium
consists of a square lattice of cylindrical tubes oriented at
$45^\circ$ to the longest side of the lattice [\fig{model}(a)]. Four
tubes meet at each intersection where we put a node having no
volume. The disorder is introduced by (1) assigning the tubes a radius
$r$ chosen at random inside a defined interval or (2) moving the
intersections a randomly chosen distance away from their initial
positions. In (1) all tubes have equal length $d$ but different
$r$. (2) results in a distorted square lattice giving the tubes
different lengths. Here $r=d/2\alpha$ where $\alpha$ is the aspect
ratio between the tube length and its radius.

The tubes are initially filled with a wetting fluid of viscosity
$\mu_w$ and a nonwetting fluid of viscosity $\mu_{nw}\ge\mu_w$, is
injected at constant injection rate $Q$ along the bottom row
(inlet). The viscosity ratio $M$ is defined as $M\equiv
\mu_{nw}/\mu_w$. The wetting fluid is displaced and flows out along
the top row (outlet). There are periodic boundary conditions in the
orthogonal direction. The fluids are assumed immiscible, hence an
interface (a meniscus) is located where the fluids meet in the
tubes. The capillary pressure $p_c$ of a meniscus is given by
$p_c=(2\gamma/r)\left[1-\cos(2\pi x/d)\right]$. The first term is
Young-Laplace law for a cylindrical tube when perfect wetting is
assumed and in the second term $x$ is the position of the meniscus in
the tube ($0\le x \le d$). Thus, with respect to the capillary
pressure we treat the tubes as if they were hourglass shaped with
effective radii following a smooth function.  By letting $p_c$ vary as
above, we include the effect of local readjustments of the menisci at
pore level~\cite{Aker98-1} which is important for the description of
burst dynamics~\cite{Hain30_Maloy96}.  The detailed modeling of $p_c$
costs computation time, but is necessary in order to properly simulate
the capillary pressure behavior along the front.

The volume flux $q_{ij}$ through a tube between the $i$th and the
$j$th node is given by Washburn equation~\cite{Wash21}:
$q_{ij}=-(\sigma_{ij} k_{ij}/\mu_{ij})(p_j-p_i-p_{c,ij})/d_{ij}$.
Here $k_{ij}$ is the permeability of the tube, $\sigma_{ij}$ is the
average cross section of the tube, $p_i$ and $p_j$ is the pressures at
node $i$ and $j$ respectively, and $p_{c,ij}$ is the sum of the
capillary pressures of the menisci inside the tube. A tube partially
filled with both liquids, is allowed to contain one or two
menisci. Furthermore, $\mu_{ij}$ denotes the effective viscosity given
by the sum of the volume fractions of each fluid inside the tube
multiplied by their respective viscosities. Inserting the above
equation for $q_{ij}$ into Kirchhoff equations at every node (volume
flux conservation), $\sum_j q_{ij}=0$, constitutes a set of linear
equations which are to be solved for $p_i$. The set of equations is
solved by using the Conjugate Gradient method with the constraint that
$Q$ is held fixed.  See Refs.~\cite{Aker98-1,Aker98-2} for details on
the numerical scheme updating the menisci and solving $p_i$.

The front between the two phases is detected by running a
Hoshen-Kopelman algorithm~\cite{Stauf92} on the lattice.  The front
width is defined as the standard deviation of the distances between
each meniscus along the front and the average front position in the
direction of the displacement. $\Delta P_c$ as function of $\Delta h$
is calculated by taking the mean of the capillary pressure differences
between all pairs of menisci separated a height $\Delta h$
along the front.  The capillary pressure difference between a pair of
menisci is calculated by taking the capillary pressure of the
meniscus closest to the inlet minus the capillary pressure of the
meniscus closest to the outlet [\fig{model}(a)].

Figure~\ref{fig:pc-m100} shows $\Delta P_c$ as function of $\Delta h$
for simulations performed at three different $C_a$'s with $M=100$ or
$1$. The simulations with $M=100$ were performed on a $25\times 35$
nodes lattice with $\mu_{nw}=10$ P, $\mu_w=0.10$ P, and $\gamma=30$
dyn/cm. The disorder was introduced by choosing the tube radii at
random in the interval $0.05d\le r_{ij}\le d$. The tube length was
$d=0.1$ cm. The simulations with $M=1$ were performed on a distorted
lattice of $40\times 60$ nodes where $0.02\ \mbox{cm}\le d_{ij}\le
0.18$ cm and $r_{ij}=d_{ij}/2\alpha$ with $\alpha=1.25$. Here
$\mu_{nw}=\mu_w=0.5$ P. To obtain reliable average quantities we did
10--30 simulations at each $C_a$ with different sets of random
$r_{ij}$ or $d_{ij}$.

From \fig{pc-m100} we observe that $\Delta P_c$ increases roughly
linearly as function of $\Delta h$.  At lowest $C_a$ no clear
stabilization of the front was observed due to the finite size of the
system.  At higher $C_a$ the viscous gradient stabilizes the
front. The gradient causes the capillary pressure of the menisci
closest to the inlet to exceed the capillary pressure of the menisci
lying in the uppermost part. Thus, the menisci closest to the inlet
will more easily penetrate a narrow tube compared to menisci further
down stream.  This will eventually stabilize the front.

To save computation time and thereby be able to study $\Delta P_c$ on
larger lattices in the small $C_a$ regime, we have generated bond
invasion percolation (IP) patterns with trapping on lattices of
$200\times 300$ nodes.  The IP patterns were generated on the bonds in
a square lattice with the bonds oriented diagonally at
$45^\circ$. Hence, the bonds correspond to the tubes in our network
model.  Each bond was assigned a random number $f_{ij}$ in the
interval $[0,1]$. A small stabilizing gradient $g=0.05$ was applied,
giving an occupation threshold $t_{ij}$ of every bond:
$t_{ij}=f_{ij}+gh_{ij}$~\cite{Wilk86,Birov91}.  Here $h_{ij}$ denotes
the height of the bond above the bottom row. The occupation of bonds
started at the bottom row, and the next bond to be occupied was always
the bond with the lowest threshold value from the set of empty bonds
along the invasion front. The generated IP patterns are similar to the
site-bond IP patterns in~\cite{Sahimi98} and we assume they are
statistical equal to structures that would have been obtained in a
corresponding complete displacement simulation.

When the IP patterns became well developed with trapped (wetting)
clusters of sizes between the bond length and the front width, the
tubes in our network model were filled with nonwetting and wetting
fluid according to occupied and empty bonds in the IP lattice.
Moreover, the radii $r_{ij}$ of the tubes were mapped to the random
numbers $f_{ij}$ of the bonds as
$r_{ij}=[0.05+0.95(1-f_{ij})]d$. Thus, $0.05d\le r_{ij} \le d$ and we
set the tube length $d=0.1\,\mbox{cm}$. Note that $r_{ij}$ is mapped
to $1-f_{ij}$ because in our IP algorithm the next bond to be invaded
is the one with the lowest threshold value, opposite to the network
model, where the widest tube will be invaded first.

After the initiation of the tube network was completed, the network
model was started and the simulations were run a limited number of
time steps before it was stopped. The number of time steps where
chosen sufficiently large to let the menisci along the front adjust
according to the viscous pressure set up by the injection rate.  
%The injection rate $Q$ was chosen in correspondence to $g$.

Totally, we generated four IP patterns with different sets of $f_{ij}$
and every pattern was loaded into the network model.  The
result of the calculated $\Delta P_c$ versus $\Delta h$ is shown in
\fig{pcd-ip} for $C_a=\exf{9.5}{-5}$ and $M=100$.  If we assume a
power law $\Delta P_c\propto \Delta h^\kappa$, we find $\kappa =
1.0\pm 0.1$. The slope of the straight line in \fig{pcd-ip} is 1.0.
We have also calculated $\Delta P_c$ for $C_a=\exf{2}{-6}$ with $M=1$
and $M=100$ by using one of the generated IP patterns. The result of
those simulations is consistent with \fig{pcd-ip}.

Wilkinson~\cite{Wilk86} was the first to use percolation theory to
deduce a power law between $w_s$ and $C_a$ when only viscous forces
stabilize the front. In 3D, where trapping of wetting fluid is assumed
to be of little importance, he suggested $w_s\propto {C_a}^{-\alpha}$
and $\alpha=\nu/(1+t-\beta+\nu)$. Here $t$ is the conductivity
exponent and $\beta$ is the order parameter exponent in percolation.
Blunt {\em et al.}\ \cite{Blunt92} used a similar approach, however,
they found $\alpha = \nu/(1+t+\nu)$ in 3D. This is identical to the
result of Lenormand~\cite{Len89} discussing limits of fractal patterns
between capillary fingering and stable displacement in 2D porous
media. Blunt {\em et al.} also deduced a scaling relation for the
pressure drop $\Delta P_{nw}$ across a height difference $\Delta h$ in
the nonwetting phase of the front and found $\Delta
P_{nw}\propto\Delta h^{t/\nu + 1}$. Later on, Xu {\em et al.}\
\cite{Xu98} used the arguments of Gouyet {\em et al.}~\cite{Sap88} and
Wilkinson~\cite{Wilk86} to show that $\Delta P_{nw}\propto {\Delta
h}^{t/\nu+d_{\text{E}}-1-\beta/\nu}$, where $d_{\text{E}}$ is the
Euclidean dimension of the space in which the front is embedded. They
also argued that $\Delta P_c=\Delta P_{nw}-\Delta P_{w}$ where $\Delta
P_w$ denoting the pressure drop in the wetting phase of the front, is
linearly dependent on $\Delta h$ due to the compact displaced fluid
[see \fig{model}(a)]. Thus, the result of Xu {\em et al.} would
in 2D predict $\Delta P_c\propto\Delta h^{1.9}$ where we have used $t=1.3$,
$\nu=4/3$, $\beta=5/36$, and $d_{\text{E}}=2$.  Our
simulations give $\Delta P_c\propto\Delta h^\kappa$ and $\kappa =
1.0\pm 0.1$. Below we present an alternative view on the displacement
pattern from those first suggested by Wilkinson. The alternative view
is based upon the loopless nonwetting strands and is supported by our
numerical result.

The simulated displacement patterns show that the nonwetting fluid
contains no closed loops [\fig{model}(a)] because wetting fluid may be
trapped in single tubes, due to volume conservation [\fig{model}(b)].
Because of fluid trapping in single tubes, the invading fluid flows in
separate strands that cannot coalesce. We note that the definition in
in \fig{model}(b) can be easily generalized to 3D~\cite{Porto97},
since increasing the coordination number of the lattice does not
change the trapping rule. Therefore, we expect loopless patterns to
develop in 3D lattices and our arguments that we present below should
apply there too. We also note that trapping of wetting fluid is more
difficult in real porous media due to a more complex topology of pores
and throats there. Loopless IP patterns have earlier been observed in
Refs.~\cite{Sahimi98,Yortsos97,Cieplak96}.

%also pointed out for a site-bond IP algorithm with trapping in
%Refs.~\cite{Sahimi98,Yortsos97} and for a loopless IP algorithm in
%Ref.~\cite{Cieplak96}.

From \fig{model}(a) we may separate the displacement pattern into two
parts: one consisting of the frontal region continuously covering new
tubes, and the other consisting of the more static structure behind
the front. The frontal region is supplied by nonwetting fluid through
strands connecting the frontal region to the inlet. When
the strands approach the frontal region they are more likely to split.
Since we are dealing with a square lattice, a splitting strand may
create either two or three new strands.  As the strands proceed
further into the frontal region they split again and again and
eventually they cover the frontal region completely [see
\fig{model}(a)].

On IP patterns without loops~\cite{Sahimi98,Porto97,Cieplak96} the
length $l$ of the minimum path between two points separated an
Euclidean distance $R$ scales like $l\propto R^{D_s}$ where $D_s$ is
the fractal dimension of the shortest path. We assume that the
displacement pattern of the frontal region for length less than the
correlation length (in our case $w_s$) is statistically equal to IP
patterns in~\cite{Sahimi98}.  Therefore, the length of a strand in the
frontal region is proportional to $\Delta h^{D_s}$ when $\Delta h$
is less than $w_s$. If we assume that on the average every tube in the
lattice has same mobility ($k_{ij}/\mu_{ij}$), this causes the fluid
pressure within a single strand to drop like $\Delta h^{D_s}$ as long
as the strand does not split. When the strand splits volume conservation
causes the volume fluxes through the new strands to be less than the 
flux in the strand before it splits. Hence, following a path where
strands split will cause the pressure to drop as $\Delta
h^\kappa$ where $\kappa\le D_s$. 

From the above arguments we conclude that the pressure drop $\Delta
P_{nw}$, in the nonwetting phase of the frontal region (that is the
strands) should scale as $\Delta P_{nw}\propto\Delta h^\kappa$ where
$\kappa\le D_s$. In 2D two different values for $D_s$ have been
reported: $D_s=1.22$~\cite{Porto97,Cieplak96} and
$D_s=1.14$~\cite{Sahimi98}. Both values are consistent with our simulations
finding $\kappa = 1.0\pm 0.1$. 

The evidence that $\kappa\simeq 1.0$ may influence the scaling of
$w_s$ as function of $C_a$. At low $C_a$ simulations show that
$\Delta\widehat{P}_c\propto C_a\Delta h^{1.0}$~\cite{Aker00-1}. Here
$\Delta\widehat{P}_c$ denotes the capillary pressure difference when
the front is stationary. That means, $\Delta\widehat{P}_c$ excludes
situations where nonwetting fluid rapidly invades new tubes due to
local instabilities.  At sufficiently low $C_a$ the displacement can
be mapped to percolation giving $\Delta\widehat{P}_c\propto
f-f_c\propto\xi^{-1/\nu}$~\cite{Wilk86,Birov91,Sap88}. Here $f$ is the
occupation probability of the bonds, $f_c$ is the percolation
threshold, and $\xi\propto w_s$ is the correlation length. By
combining the above relations, we obtain $w_s\propto {C_a}^{-\alpha}$
where $\alpha=\nu/(1+\nu\kappa)$. In 2D $\nu =4/3$ and inserting
$\kappa=1.0$ gives $\alpha\approx 0.57$. At high $C_a$ we expect a
crossover to another type of behavior since it is not clear if the
mapping to percolation~\cite{Wilk86,Birov91,Sap88} is valid there.  We
note that Wilkinson's result~\cite{Wilk86} gives $\alpha\approx0.38$
in 2D.

In summary we conclude that $\Delta P_c\propto\Delta h^\kappa$ where
our simulations gives $\kappa=1.0\pm 0.1$. By describing the
displacement structure in terms of loopless
strands~\cite{Sahimi98,Cieplak96} we have argued that $\kappa\le D_s$,
where $D_s$ is the fractal dimension of the shortest path between two
points on IP patterns without loops. In 2D two values of $D_s$ has
been reported ($1.14$~\cite{Sahimi98} and
$1.22$~\cite{Porto97,Cieplak96}) and both are consistent with our
numerical result $\kappa\simeq 1.0$. We conclude that earlier
suggested theories~\cite{Wilk86,Len89,Blunt92,Xu98} are not compatible
in situations where a loopless pattern with nonwetting strands
dominate the displacement. We have also shown that $\alpha$ in
$w_s\propto {C_a}^{-\alpha}$, may be influenced by the evidence that
$\kappa\le D_s$. Work is in progress to investigate our arguments in 3D 
and the effect of loops on $\kappa$.

The authors thank J.\ Feder and E.\ G.\ Flekk\o y for valuable comments.
The work is supported by the Norwegian Research Council (NFR) through a 
``SUP'' program and we acknowledge them for a grant of computer time.

%\bibliographystyle{$HOME/TEX/bst/prl_roman} 
%\bibliography{$HOME/TEX/bib/porous}

\end{document}